\documentclass[pra,twocolumn,a4paper,showpacs,superscriptaddress]{revtex4}

 \usepackage{latexsym}
 \usepackage{amsmath}
 \usepackage{amsfonts}
 \usepackage{graphicx}
 \usepackage{amssymb}
 \usepackage{subfigure}
 \usepackage{color}
 \usepackage{setspace}
 
 \newcommand{\ket}[1]{\ensuremath{|#1\rangle}}
 \newcommand{\bra}[1]{\ensuremath{\langle #1 |}}

 \newcommand{\bc}{\begin{center}}
 \newcommand{\ec}{\end{center}}
 \newcommand{\mf}[1]{\boldsymbol{#1}}
 
 \newcommand{\ii}{i}

\begin{document}

\title{Subluminal light pulse propagation in a $\Lambda$-type atom-molecule coupled system }

\author{Sandeep \surname{Sharma}}
\email{sandeep.sharma@iitg.ernet.in}
\affiliation{Department of Physics, Indian Institute of Technology Guwahati,
Guwahati- 781039, Assam, India}

\author{Bimalendu \surname{Deb}}
\email{msbd@iacs.res.in}
\affiliation{Department of Materials Science, Indian Association for the Cultivation of Science,
Jadavpur, Kolkata- 700032, India}

\author{Tarak N. \surname{Dey}}
\email{tarak.dey@gmail.com}
\affiliation{Department of Physics, Indian Institute of Technology Guwahati,
Guwahati- 781039, Assam, India}
\date{\today}

\begin{abstract}
We show theoretically the presence of electromagnetically induced transparency (EIT) like feature in a $\Lambda$-type atom-molecule coupled system. 
We demonstrate that EIT in such a systems has a sharp dispersion characteristic which is useful to slow down light propagation in a medium.
The results of our model calculations show that, it is possible to achieve a time delay of the order of 1 ms for a probe field propagating a distance of 1 cm through a medium of ultracold diatomic molecules coupled to the continuum of atom-atom collisional states with a control field. 
This enables one to reduce the group velocity of the light much below 100 m s$^{-1}$ with more than 25$\%$ transmission.
\end{abstract}
\maketitle
\section{Introduction}
Electromagnetically induced transparency (EIT) \cite{Boller, Field} is an effect by which a medium is rendered transparent to a weak (probe) light field. 
This happens due to the manipulation of medium's optical properties by a strong (control) light field. 
EIT has applications in many areas such as high precision spectroscopy \cite{Scully, Lukin}, atomic clocks \cite{Kitching, Knappe}, slow light \cite{Kasapi, Schmidt, Hau}, fast light \cite{Wang, Dogariu}, \textit{etc}. 
Harris \textit{et al.} have shown that the presence of a strong control field leads to a narrow transparency window in the absorption profile of a probe \cite{Harris1, Harris2}.
Further, it is noticed that in the limit of absorption going to zero, the dispersion profile shows a sharp variation over the spectral range of the window. 
A positive slope of the dispersion in this narrow spectral region leads to slow light \cite{Boyd, Wang1}, while a negative slope makes light faster \cite{Wang}.
Currently, applications of slow light is hindered due to a practical difficulty. 
This stems from the fact that, a narrow width of the transparency window gives rise to a large time-delay and high absorption, whereas a broad width will lead to a very small time-delay and less absorption. 
Hence, to optimize the outcome of any slow-light experiment, a quantity termed as Delay-bandwidth product (DBP) has been introduced by Tidstrom \textit{et al.} \cite{Tidstrom}. 
DBP is the product of time-delay and the spectral width of transparency window. 
It is used to characterize the performance of slow light devices.
 
The main difficulty in producing slow light is to get a delay that is comparable to the inverse of the probe pulse width without causing any distortion to the pulse. 
We show that an atom-molecule coupled system at low energy is useful to overcome this difficulty. In general, it is difficult to generate slow light in a molecular system because of the complex level structure of the molecules unlike that in an atomic gas. 
Recent advances in the creation of ultracold diatomic molecules in a specific rovibrational level by associating ultracold atoms open up prospects for using molecular states for coherent control of light propagation. 
One of the prominent methods to associate cold atoms into molecules is photoassociation (PA) \cite{Thorsheim, Weiner, Jones}. 
There is another method of association of ultracold atoms to form a class of exotic molecules known as Feshbach molecules \cite{Kohler, Chin_3}. 
This is based on three-body interactions in the presence of a magnetic field tunable Feshbach resonance. 
In this work we do not consider Feshbach molecules and focus only on usual molecular states.

\begin{figure}[b!]
\includegraphics[width=0.9\columnwidth]{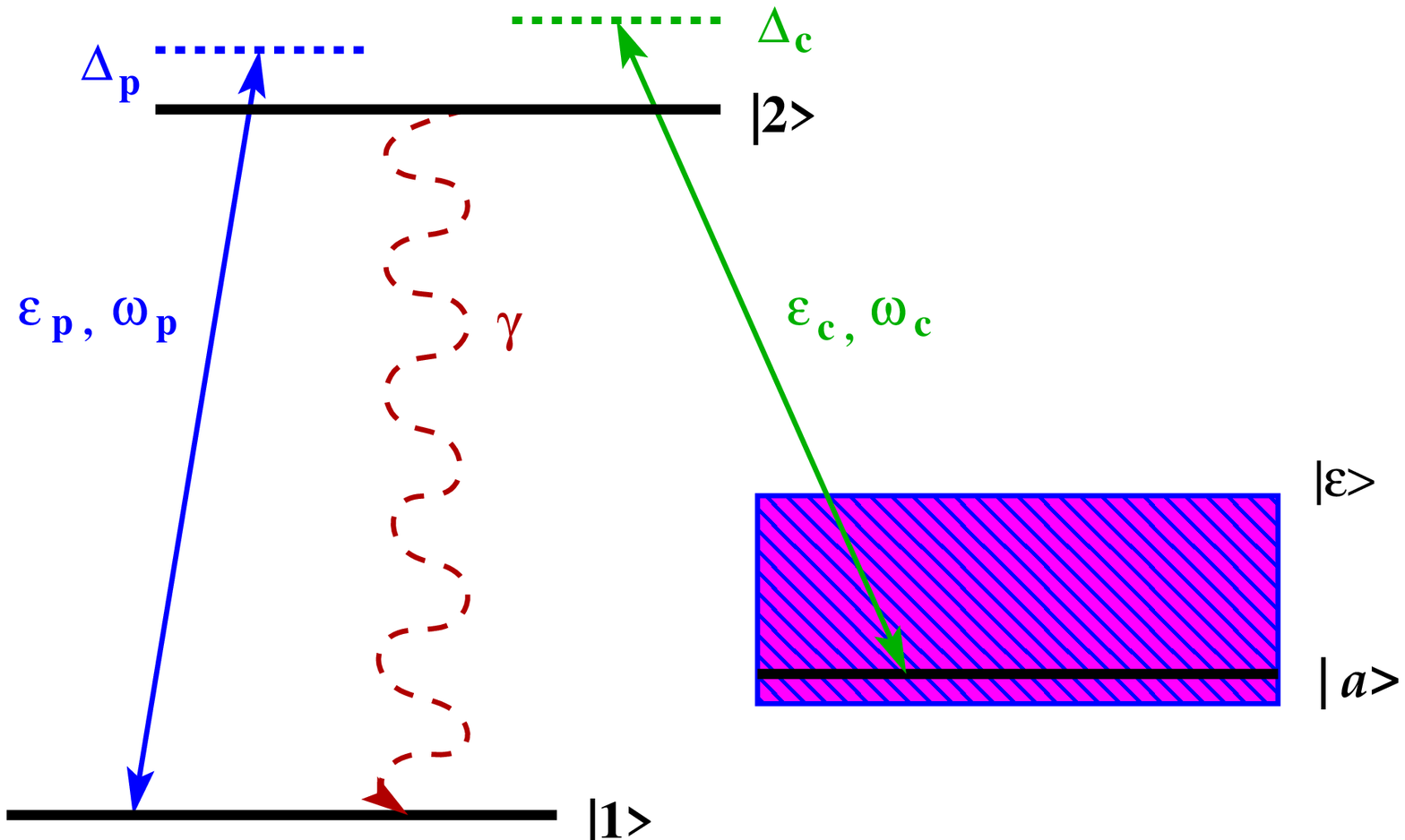}
\caption{\label{fig:Fig1} (Color online) 
 Schematic diagram of the model system. The initial state $\ket{1}$ is coupled to the state $\ket{2}$ via a probe field with frequency $\omega_p$ and the state $\ket{2}$ is coupled to the continuum $\ket{\varepsilon}$ via a control field with frequency $\omega_c$. $\ket{a}$ is a quasi-bound state coupled to the continuum via interaction $V(\varepsilon)$. $\gamma$ is the spontaneous linewidth for $\ket{2}\leftrightarrow\ket{1}$ transition.}
\end{figure}
We consider a model consisting of 3 diatomic molecular bound states and the collisional continuum of two ground state atoms in a pump-probe $\Lambda$-type configuration as shown in Fig.\ref{fig:Fig1}. 
There are two approaches to solve the problem of a continuum-bound coupled system, one is quasi-continuum \cite{Vardi, Li, Mackie} and the other is structured continuum approach \cite{Kuznetsova}.
The latter approach was introduced more than 50 years ago by Fano, who showed that a system coupled to a continuum can lead to quantum interference effects in spectral profile\cite{Fano}. 
In his work on quantum mechanical study of autoionizing states, he showed that quantum interference can originate from the interaction of a bound state with a continuum.
Further, he explicitly showed that the photoionization spectrum of a bound state coupled to both autoionizing state as well as continuum can be made zero in a certain parameter regime. This vanishing of photoionization is because of destructive interference between two different excitation pathways. 
In the case of autoionization, the two pathways are direct excitation from bound to continuum and indirect excitation from bound to continuum via autoionization state.
It was theoretically predicted by Paspalakis \textit{et al.} \cite{Paspalakis}, that Fano interference arising form the interaction of an autoionizing medium and a short pulse can lead to soliton-like pulse propagation. 
Structured continuum and resulting quantum interference effects similar to those in autoionization also arise in PA in the presence of magnetic Feshbach resonance (MFR) \cite{Deb, Deb1}.
Early experimental works showing suppression of photoassociative atom loss due to tunable MFR induced continuum structure were carried out by Vuletic and coworkers \cite{Vuletic} and Chin \textit{et al.} \cite{Chin1}.
Moal \textit{et al.} \cite{Moal} have shown Fano profile and dark resonance in two-photon PA.
Junker \textit{et al.} have observed Fano-like asymmetric spectral profile in PA in the presence of MFR\cite{Junker}. 
Stimulated Raman Adiabatic Passage (STIRAP) with  structured continuum have been used for an efficient formation of cold molecules\cite{Bergmann, Cubel}. 
Recently, Dumke \textit{et al.} \cite{Dumke}, and Winkler \textit{et al.} \cite{Winkler} have showed the presence of EIT like feature in atom-molecule coupled systems.  

In this paper we study light propagation through an atom-molecule resonantly interacting medium. We assume that the system is initially prepared in the ground molecular state  by Raman PA \cite{Winkler, Aikawa} or by MFR in the presence of radio-frequency field \cite{Tscherbul}. 
Further, one of the states in the system is a Feshbach resonant collisional continuum as schematically shown in Fig.\ref{fig:Fig1}. 
We then use Fano theory to reduce the composite system of bound state interacting with the continuum into a single broadened discrete state. 
This system under the presence of strong control field shows EIT like feature. We find that with the decrease in the Feshbach resonance width, the width of the transparency window becomes very narrow and the absorption goes to zero. 
We exploit these features to reduce significantly the group velocity of the propagating weak pulse and to increase the output transmission efficiency.

The paper is organized as follows. In Sec.II A, we introduce our model system. 
In Sec.II B, we study the dynamics of the system and describe the linear susceptibility of the probe. 
In Sec.II C, we analyze the linear response of the medium in steady state for various cases. 
In Sec.II D, we numerically study the pulse propagation effects for both Gaussian and hyperbolic secant pulses and discuss how to optimize DBP for this system. 
Finally in Sec.III we summarize our results and make comments on the applications of the predicted slow light effects. 
 
\section{Theoretical Formulations}
\subsection{Model}
We consider a $\Lambda$-type atom-molecule coupled system. Our model is schematically shown in Fig.\ref{fig:Fig1} with $\ket{1}$ and $\ket{2}$ being two molecular bound states, $\ket{a}$ representing a quasibound or predissociating state embedded in the continuum $\ket{\varepsilon}$ of scattering states of two ground state atoms. The excited rovibrational state $\ket{2}$ is coupled to the rovibrational state $\ket{1}$ in an electronic molecular ground state by a weak probe field with frequency $\omega_p$ and to the continuum $\ket{\varepsilon}$ by a strong control field with frequency $\omega_c$, respectively. The two fields are given by
\begin{equation}
\label{Equation1}
 {\vec{E}_j}(z,t)= \hat{e}_{j}\mathcal{E}_{j}(z,t)~e^{- i\left(\omega_j t-
 k_j z\right )} + {c.c.}\,,
\end{equation}
where $\mathcal{E}_{j}(z,t)$ is the slowly varying electric field envelope, $\hat{e}_{j}$ is the unit polarization vector, $\omega_j$ is the laser field frequency and $k_j$ is
the wave number of the field. 
The index $j\in \{p,c\}$ denotes the probe ($p$) and control ($c$) field, respectively. 
The proposed scheme may be realized in a degenerate or near-degenerate molecular gas prepared in a single vibrational and rotational state in electronic ground-state potential.  
Recently, the production of ultracold KRb molecular gas in high phase space density has been experimentally demonstrated \cite{Ni, Aikawa}. 
For numerical illustration of our work, we take the singlet electronic ground state $X^{1}\Sigma$ of KRb with the vibrational level $v=0$ as $\ket{1}$. 
The electronic excited state $2^{3}\Sigma$ with the vibrational level $v^{\prime}=23$ is assumed to be $\ket{2}$ which is coupled to $\ket{1}$ and $\ket{\varepsilon}$ by two coherent fields, namely a weak probe field with wavelength 690 nm and a control field with wavelength 970 nm, respectively. 
The existence of a quasi-bound state $\ket{a}$ is assumed to model a MFR under two-channel approximation, where one of the channels that can support $\ket{a}$ is closed, and the other channel is open implying that the continuum of scattering states $\ket{\varepsilon}$ belong to open channel. 
The dipole moment of the transition between the ground and excited electronic states is $0.012 ea_{0}$. 
The radiative decay rate $\gamma$ is taken to be $1.8$ MHz.
\subsection{Theory}
According to Fano theory, when the state $\ket{a}$ with energy $E_{a}$ is embedded into the flat continuum $\ket{\varepsilon}$ via an interaction $V(\varepsilon)$( generally collisional interaction), a shift results in the energy of the state $\ket{a}$ leading to its dressing by the continuum. 
The flat continuum states $\ket{\varepsilon}$ also gets modified by the interaction $V(\varepsilon)$ to produce the dressed continuum states $\ket{\epsilon}$. 
The bound to dressed continuum dipole moment $d_{2\epsilon}$ is related to bound to flat continuum dipole moment $d_{2\varepsilon}$ as
\begin{subequations}
\begin{align}
\label{Equation2}
d_{2\epsilon} =& \frac{\epsilon-E_{a}+q\Gamma}{\sqrt{(\epsilon-E_{a})^{2}+(\Gamma)^{2}}}d_{2\varepsilon}\,, \\
d_{2\varepsilon} =& \frac{\sqrt{2}\mf{\bar{d}}}{q\sqrt{\pi \Gamma}}\,,
\end{align}
\end{subequations}
where $\mf{\bar{d}}$ is the matrix element of dipole moment between the ground and excited molecular electronic states, and $q$ is the well-known Fano asymmetry parameter expressed as
\begin{subequations}
\begin{align}
\label{Equation3}
q =& \frac{\bra{b}d\ket{2}}{\pi \bra{2}d\ket{\varepsilon} \bra{\varepsilon}d\ket{a}}\,, \\
\ket{b} =& \ket{a} + P \int \frac{V(\varepsilon)}{\epsilon-\varepsilon}\ket{\varepsilon}\, d\varepsilon \,,
\end{align}
\end{subequations}
and $\Gamma$ is the half-width of the structure in the continuum $\ket{\varepsilon}$ due to the embedding of the discrete state $\ket{a}$. 
In the case of Feshbach resonance, $2 \Gamma$ can be identified with the width of the resonance.
In the dressed continuum picture, the wave-function of the system is a coherent superposition of the two bound states $\ket{1}$, $\ket{2}$ and the dressed continuum $\ket{\epsilon}$ as given by
\begin{equation}
\label{Equation4}
\ket{\psi(t)} = C_{1}(t)\ket{1} + C_{2}(t)\ket{2} + \int C_{\epsilon}(t)\ket{\epsilon}\, d\epsilon \,,
\end{equation}
where $C_{1}$, $C_{2}$ and $C_{\epsilon}$ are the superposition coefficients. The corresponding density matrix is expressed as
\begin{equation}
\label{Equation5}
\rho = \ket{\psi(t)}\bra{\psi(t)}  \,.
\end{equation}
The time-dependent Hamiltonian of the system under the electric dipole approximation can be written as
\begin{subequations}
\label{Equation6}
\begin{align}
H =& H_0 + H_I\,,\\
H_0 =& \hbar\omega_{1}\ket{1}\bra{1} + \hbar\omega_{2} \ket{2}\bra{2} +
\int \epsilon \ket{\epsilon}\bra{\epsilon}\, d\epsilon \,,\\
H_I =& -\ket{1}\bra{2} \mf{d}_{12} [\mathcal{E}_{p}e^{- i\omega_p t} + c.c.]  \nonumber\\
&  - \ket{2}\bra{1} \mf{d}_{21} [\mathcal{E}_{p}e^{- i\omega_p t} + c.c.] \nonumber\\   
&  - \int \ket{\epsilon}\bra{2} \mf{d}_{\epsilon2} [\mathcal{E}_{c}e^{- i\omega_c t} + c.c.]\, d\epsilon \nonumber\\ 
&  - \int \ket{2}\bra{\epsilon} \mf{d}_{2\epsilon} [\mathcal{E}_{c}e^{- i\omega_c t} + c.c.]\, d\epsilon \,.
\end{align}
\end{subequations}
where $\mf{d}_{ij} = \bra{i}d.\hat{e}\ket{j}$ are dipole moments corresponding to $\ket{i}\leftrightarrow\ket{j}$ transition. 
The free-bound stimulated linewidth due to control field is given by $\Omega_{c} = \frac{\left\vert\mf{d}_{\varepsilon 2} \mathcal{E}_{c}\right\vert^{2}}{\hbar}$. 
The dynamics of the atom-molecule coupled system is governed by the following master equation
\begin{equation}
\label{Equation7}
 \dot{\rho} =\frac{i}{\hbar} [H,\rho] - \mathcal{L}\rho   \,.
\end{equation}
The term $\mathcal{L}\rho$ is Liouville operator which represents all the incoherent processes and is given by
\begin{equation}
\label{Equation8}
\mathcal{L}\rho = \gamma\left(\rho\ket{2}\bra{2}-2\rho_{22}\ket{1}\bra{1}+\ket{2}\bra{2}\rho\right)   \,.
\end{equation}
where $\gamma$ is the spontaneous decay rate of state $\ket{2}$ to $\ket{1}$. 
Note that, since the transition $\ket{2}\leftrightarrow\ket{\epsilon}$ is driven by a strong control field we have neglected bound to continuum spontaneous emission.
In the weak probe field limit, the dynamics of atomic coherences for the $\Lambda$ system are governed by
\begin{subequations}
\begin{align}
\label{Equation9}
\ii \hbar \dot{\rho}_{21} =& \hbar(\omega_{2}-\omega_{1}) \rho_{21} - \mf{d}_{21}[\mathcal{E}_{p}e^{- i\omega_p t} + c.c.] \nonumber \\
&  - \int \rho_{\epsilon1} \mf{d}_{2\epsilon} [\mathcal{E}_{c}e^{- i\omega_c t} + c.c.]\, d\epsilon - \ii \hbar \gamma \rho_{21} \,. \\
\ii \hbar \dot{\rho}_{\epsilon1} =& (\epsilon-\hbar\omega_{1}) \rho_{\epsilon1}- \rho_{21}\mf{d}_{\epsilon2} [\mathcal{E}_{c}e^{- i\omega_c t} + c.c.] \,.
\end{align}
\end{subequations}  
Under the unitary transformations $\sigma_{21} = \rho_{21}e^{i\omega_p t}$, $\sigma_{\epsilon1} = \rho_{\epsilon1}e^{i(\omega_p-\omega_c) t}$ and rotating wave approximation Eq.~(\ref{Equation9}-9b) reduce to
\begin{subequations}
\begin{align}
\label{Equation10}
\ii \hbar \dot{\sigma}_{21} =& -( \hbar \Delta_{p}+ \ii \hbar \gamma) \sigma_{21} - \mf{d}_{21}\mathcal{E}_{p} \nonumber\\
& - \int \rho_{\epsilon1} \mf{d}_{2\epsilon} \mathcal{E}_{c}\, d\epsilon \,. \\ 
\ii \hbar \dot{\sigma}_{\epsilon1} =& (\epsilon- \hbar\omega_{1}- \hbar\omega_{p}+ \hbar\omega_{c})\sigma_{\epsilon1}-\sigma_{21}\mf{d}_{\epsilon2} \mathcal{E}^{*}_{c}   \,.
\end{align}
\end{subequations}  
Next we use these density matrix equation to find out the probe susceptibility in the steady state limit. We then study the dependency of probe susceptibility on $\Omega_{c}$ and resonance width $\Gamma$.
\subsection{Susceptibility of the medium}
In the steady state limit, the probe susceptibility $\chi$ of the medium at frequency $\omega_{p}$ can be determined by the atomic coherence $\sigma_{21}$.
From Eq.~(\ref{Equation10}) we have
\begin{subequations}
\begin{align}
\label{Equation11}
\sigma_{21} =&   \frac{-\mf{d}_{21}\mathcal{E}_{p}}{\hbar \Delta_{p}+ \ii \hbar \gamma + \mathcal{E}^{2}_{c}\beta}\,, \\ \nonumber
\beta =&   \int\frac{\mf{d}_{2\epsilon}\mf{d}_{\epsilon2}}{\epsilon+ \hbar\omega_{c}- \hbar\omega_{1}- \hbar\omega_{p}}\, d\epsilon\,.
\end{align}
\end{subequations}
The polarization of the medium in terms of atomic coherence is given by
\begin{figure}[t!]
\includegraphics[width=0.7\columnwidth]{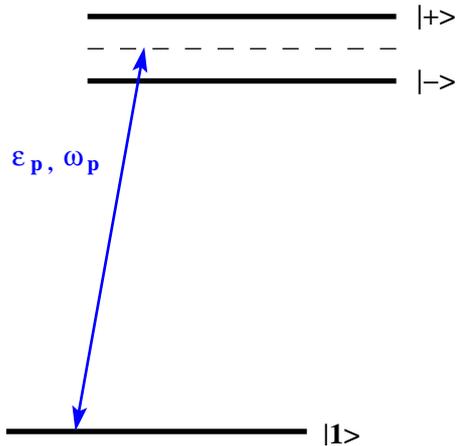}
\caption{\label{fig:Fig2} (Color online) 
Schematic diagram of the three level $\Lambda$-type system in the dressed state picture in a weak probe field limit.}
\end{figure}
\begin{equation}
\label{Equation12}
P(\omega_{p}) = \mathcal{N} \mf{d}_{12}\sigma_{21} \,.
\end{equation}
where $\mathcal{N}$ is the density of molecular medium. 

\begin{figure}[b!]
\includegraphics[width=1.\columnwidth]{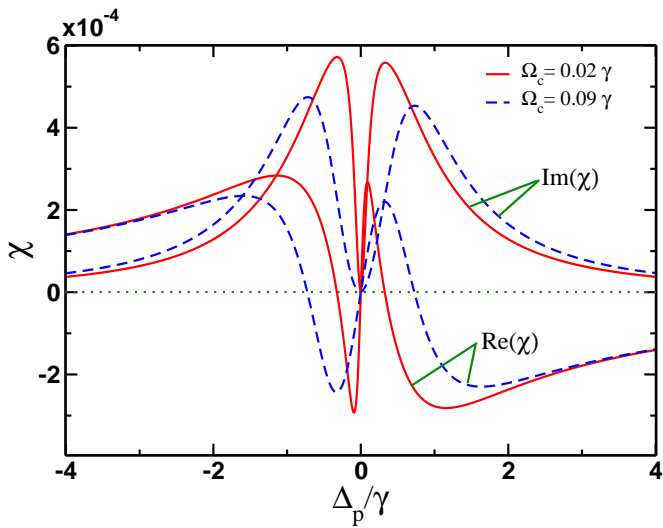}
\caption{\label{fig:Fig3} (Color online) 
The variation of the real and imaginary parts of the susceptibility are plotted as a function of $\Delta_{p}$ (in unit of $\gamma$) for different $\Omega_{c}$= 0.02 $\gamma$ (red-solid), $\Omega_{c}$= 0.09 $\gamma$ (blue-dashed). Other parameters used are $q=50$, $N = 10^{14}$ molecules/c.c., $\Gamma$ = $10^{-4} \gamma $.}
\end{figure}
Also polarization can be expressed in terms of medium susceptibility $\chi(\omega_{p})$ as
\begin{equation}
\label{Equation13}
P(\omega_{p}) = \epsilon_{0}\chi(\omega_{p})\mathcal{E}_{p} \,.
\end{equation}
Both the above Eqs.~(\ref{Equation12}) and (\ref{Equation13}) leads to the probe susceptibility as
\begin{equation}
\label{Equation14}
\chi(\omega_{p}) = -\frac{\mathcal{N}}{\epsilon_{0}}\frac{|\mf{d}_{12}|^{2}}{(\hbar \Delta_{p}+ \ii \hbar \gamma + \mathcal{E}^{2}_{c}\beta)} \,.
\end{equation}
\begin{equation}
\label{Equation15}
\beta = \lim_{\zeta \to 0} \int_{\epsilon_{th}}^{\infty}\frac{\mf{d}_{2\epsilon}\mf{d}_{\epsilon2}}{\epsilon+ \hbar\omega_{c}- \hbar\omega_{1}- \hbar\omega_{p}- \ii \zeta}\, d\epsilon \,.
\end{equation}
In order to get an analytical expression of the probe susceptibility as in Eq.~(\ref{Equation14}), we proceed by expressing bound to the dressed continuum dipole moment $\mf{d}_{2\epsilon}$ in terms of bound to bare continuum dipole moment $\mf{d}_{2\varepsilon}$ using Eq.~(\ref{Equation2}) and then substituting it in the Eq.~(\ref{Equation15}). 
Next we extend $\epsilon_{th}$ to $-\infty$ as $\mf{d}_{2\varepsilon}$ is a slowly varying function of energy near resonance, and tends to zero at large energies \textit{i.e.}, very far from the resonance. 
Hence $\mf{d}_{2\varepsilon}$ can safely be taken as constant in energy. 
This approach has been used earlier for solving bound-continuum coupled problems \cite{Kuznetsova, Fano1, Lambropoulos, Ravi, Raczynski}. 
Further, we use Cauchy's residue method to solve the integral in Eq.~(\ref{Equation15}) and arrive at
\begin{equation}
\label{Equation16}
\beta = \pi |\mf{d}_{2\varepsilon}|^{2}\left(\frac{(q-\ii)^{2}\Gamma}{(\hbar\Delta_{c}-\hbar\Delta_{p}) -\ii \Gamma}+\ii \right) \,.
\end{equation}
The behavior of the probe susceptibility as in Eq.~(\ref{Equation14}) can be explained by the dressed state analysis.
In the presence of a strong control field and a weak probe field, the states $\ket{1}$, $\ket{2}$, and $\ket{\epsilon}$ are no longer the eigenstates of the Hamiltonian in Eq.~(\ref{Equation6}). 
The new eigenstates or dressed states of the Hamiltonian of Eq.~(\ref{Equation6}) are $\ket{1}$, $\ket{+}$, and $\ket{-}$ as in Fig.\ref{fig:Fig2}. 
The dressed states $\ket{+}$, and $\ket{-}$ are separated by $\Omega_{c}$.
The right and left absorption peaks of Fig.\ref{fig:Fig3} are basically due to the transitions from the ground state $\ket{1}$ to the two dressed states $\ket{+}$ and $\ket{-}$, respectively.
However in the absence of control field, the probe field will be highly absorbed when the probe frequency $\omega_{p}$ is closely tuned to $\ket{1}\leftrightarrow\ket{2}$ transition. 
Since the spacing between the state $\ket{+}$ and state $\ket{-}$ is proportional to $\Omega_{c}$, as $\Omega_{c}$ increases the spacing between the state $\ket{+}$ and state $\ket{-}$ increases. 
Hence, the transparency window increases with the increase of $\Omega_{c}$ as can be seen from Fig.\ref{fig:Fig3}. 
It is well known that any change in the absorption profile leads to considerable changes in the dispersion of the system \cite{Harris1}. 
As $\Omega_{c}$ increases, there is a broadening in the transparency window leading to a decrease in the slope of the dispersion as shown in Fig.\ref{fig:Fig3}. 
So, by controlling the slope of the dispersion one can control the group velocity of the probe field \cite{Harris1}.
\begin{figure}[t!]
\includegraphics[width=1.\columnwidth]{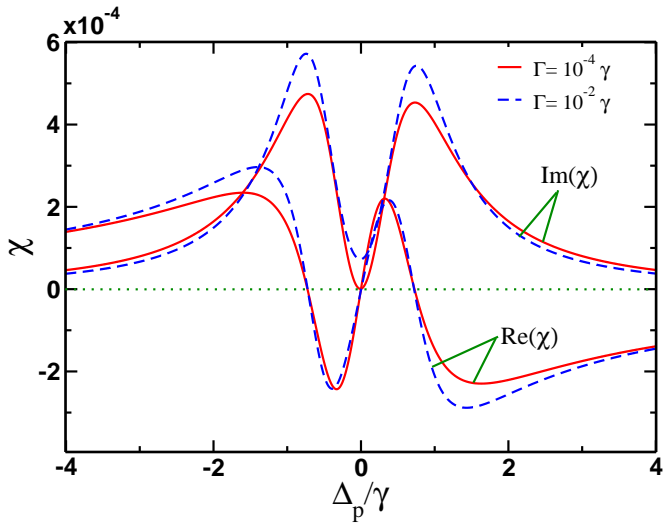}
\caption{\label{fig:Fig4} (Color online) 
 The variation of the real and imaginary parts of the susceptibility are plotted as a function of $\Delta_{p}$ (in unit of $\gamma$) for different resonance widths $\Gamma$ = $10^{-4} \gamma$ (red-solid), $\Gamma$ = $10^{-2} \gamma$ (blue-dashed). Other parameters remain same as in Fig.\ref{fig:Fig3}. except for $\Omega_{c}$= 0.09 $\gamma$.}
\end{figure}

Next, we investigate probe susceptibility for different resonance widths $\Gamma$ while keeping $\Omega_{c}$ fixed.
The aim here is to get a very narrow spectral window along with very less absorption.
Although all other parameters used for our numerical calculations correspond to KRb molecule, we have used arbitrary values of $\Gamma$ to illustrate our model.
The linewidth $\Gamma$ of magnetic Feshbach resonance for collision between K and Rb at low energy can be calculated from the parameters given in Refs. \cite{Chin_3, Simoni, Simoni1} and is estimated to be about 3.8 $\gamma$ which is much larger than what is found to be suitable for achieving sharp variation in dispersion as shown in Fig.4.
Thus, our model will be realizable for ultracold diatomic systems having very narrow Feshbach resonance line width \cite{Chin4, Chin5, Inouye, Marte} or one has to use another strong laser to have double resonances \cite{Deb2} or Autler-Towns splitting \cite{Bauer} for narrowing of resonance width.  
We can notice from Fig.\ref{fig:Fig4} that for $\Gamma$ = $10^{-4} \gamma$, the transparency window becomes very narrow and the absorption goes to zero. 
However, with the increase in $\Gamma$, the width of the absorption profile increases, because the probability of transitions from ground state $\ket{1}$ to the two dressed states $\ket{+}$ and $\ket{-}$ increases. 
This increase in the width of the absorption profile leads to a decrease in the width of the transparency window along with an increase in the absorption. 
Hence, all further studies are done with $\Gamma$ = $10^{-4} \gamma$ which is found to be most suitable value for slow light effect. 
The asymmetry shape in the EIT spectrum can be explained by the Fano parameter $q$. 
A small Fano parameter $q$ results in when $\ket{2}$ is equally bound to both the states $\ket{a}$ and the bare continuum $\ket{\varepsilon}$ \textit{i.e.}, the transition corresponding to $\ket{2}\leftrightarrow\ket{a}$ and $\ket{2}\leftrightarrow\ket{\varepsilon}$ are of equal strength. 
Hence the EIT line shape is determined by both these transitions, resulting in an asymmetric profile. 
However, this asymmetry can be removed by taking a very large Fano parameter $q$ for which, the transition $\ket{2}\leftrightarrow\ket{a}$ is very strong compared to the transition $\ket{2}\leftrightarrow\ket{\varepsilon}$. 
In this case, the line shape is solely determined by $\ket{2}\leftrightarrow\ket{a}$ transition which has Lorentzian profile and EIT spectrum tends to become symmetric.
\subsection{Propagation dynamics}
In order to see the effect of optical response on the pulse propagation dynamics,  we consider  a Gaussian probe pulse propagation \cite{Agarwal} through $\Lambda$-type atom-molecule medium of length $L$.
The envelope form of the input probe field is given by 
\begin{equation}
\label{Equation17}
E(z=0,\tau) = {\mathcal E}_{0}e^{-\frac{(\tau^{2}-\tau^{2}_{p})}{\sigma^{2}_{p}}} \,.
\end{equation}
\begin{figure}[b!]
\includegraphics[width=0.8\columnwidth]{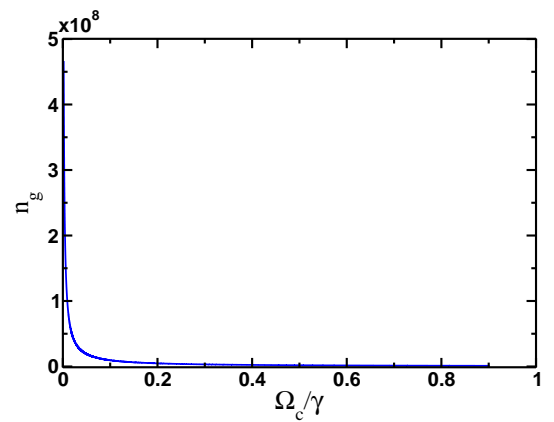}
\caption{\label{fig:Fig5} (Color online) 
The variation of group index $n_{g}$ is plotted as a function of $\Omega_{c}$ (in unit of $\gamma$). Other parameters remain same as in Fig.\ref{fig:Fig3}.}
\end{figure}
where ${\mathcal E}_0$ is the amplitude and $\sigma_p$ is the width of the input pulse. 
After traveling a distance $L$ through a medium, the output probe field has the form
\begin{figure}[t!]
\includegraphics[width=0.9\columnwidth]{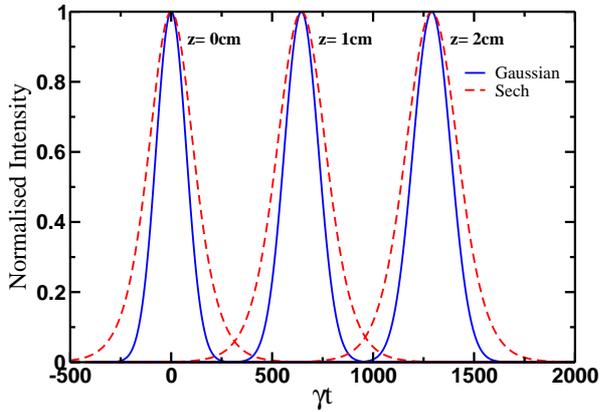}
\caption{\label{fig:Fig6} (Color online) 
 Propagation of both Gaussian pulse (blue-solid line) and hyperbolic secant pulse (red-dashed line) is shown for a spectral width of 12 KHz for different medium length. The time delay at 2 cm was $0.7$ ms and the transmission of the probe is 11.0$\%$. Other parameters remain same as in Fig.\ref{fig:Fig3}. except for $\Omega_{c}$= 0.09 $\gamma$.}
\end{figure}
\begin{subequations}
\begin{align}
\label{Equation18}
E(z=L,\tau) =& \int_{-\infty}^{\infty} E(z=0,\omega)e^{(-i(\omega t- \frac{\omega}{c}n(\omega)L)}\,d\omega \,. \\ \nonumber
n(\omega) =& 1 + 2\pi\chi(\omega) \,.
\end{align}
\end{subequations}
where $E(z=0,\omega)$ is the Fourier transform of initial input pulse $E(z=0,\tau)$ and is given by
\begin{equation}
\label{Equation19}
E(z=0,\omega) = \frac{{\mathcal E}_{0}}{\sqrt{\pi \sigma^{2}}}e^{-\frac{(\omega^{2}-\omega^{2}_{p})}{\sigma^{2}}} \,.
\end{equation}
If we assume that the refractive index varies slowly around $\omega=\omega_{p}$ then
\begin{equation}
\begin{split}
\label{Equation20}
\omega n(\omega)& = \omega_{p} n(\omega_{p})+ (\omega-\omega_{p})\frac{\partial (\omega n(\omega))}{\partial \omega}|_{\omega=\omega_{p}} \\
& + \frac{(\omega-\omega_{p})^{2}}{2}\frac{\partial^{2} (\omega n(\omega))}{\partial \omega^{2}}|_{\omega=\omega_{p}}  \,.
\end{split}
\end{equation}
substituting Eq.~(\ref{Equation19}) and Eq.~(\ref{Equation20}) in Eq.~(\ref{Equation18}), we arrive at
\begin{equation}
\begin{split}
\label{Equation21}
E(z=L,\tau)& = \frac{{\mathcal E}_{0}}{\sqrt{1-iL\kappa}} e^{(-i(\omega_{p} t- \frac{\omega_{p}}{c}n(\omega_{p})L)}\\
& e^{-\frac{\sigma^{2}}{4}\frac{(t-L/v_{g})^{2}}{1-iL\kappa}} \,.
\end{split}
\end{equation}
where $v_{g}$ is the group velocity of the pulse and is given by  
\begin{equation}
\label{Equation22}
v_{g} = \frac{c}{1+2\pi Re(\chi(\omega_{p})) + 2\pi \omega_{p}\frac{\partial Re(\chi(\omega_{p}))}{\partial \omega_{p}}}=\frac{c}{n_{g}} \,. 
\end{equation}
and the group velocity dispersion (GVD) is denoted by $\kappa$ and is defined as
\begin{equation}
\label{Equation23}
\kappa = \frac{\sigma^{2}}{2c}\frac{\partial^{2} (\omega n(\omega))}{\partial \omega^{2}}|_{\omega=\omega_{p}} \,. 
\end{equation}
For a positive $v_{g}$, the group delay is $\tau_{g}$ and is expressed as
\begin{equation}
\label{Equation24}
\tau_{g} = \left(\frac{L}{v_{g}}-\frac{L}{c}\right) \,. 
\end{equation}
One can see from the denominator of Eq.~(\ref{Equation22}) that group index $n_{g}$ depends upon the slope of the dispersion. 
The slope of the dispersion can be manipulated by changing $\Omega_{c}$. 
Hence the group index $n_{g}$ decreases with increase of $\Omega_{c}$ as shown in Fig.\ref{fig:Fig5}. 
It is evident from Eq.~(\ref{Equation21}) that real and imaginary parts of both the refractive index $n(\omega)$ and GVD play an important role to modify the amplitude and width of the incident probe pulse. 
Therefore the amplitude reduction and spreading of the pulse width leads to shape distortion along the length of the propagation. 
The main aim is to obtain distortion and absorption less pulse propagation with large group delay. 
This can be achieved by tailoring the dispersion of the medium using coherent control field.
For this purpose we study the propagation of probe pulse with two different shapes namely, a Gaussian pulse and a hyperbolic secant pulse in this medium. 
We numerically integrate the output pulse using Eq.~(\ref{Equation18}). 
Fig.\ref{fig:Fig6} shows pulse envelope for both the Gaussian and hyperbolic secant pulse at different propagation distances. 
From Fig.\ref{fig:Fig6}, we notice a considerable slowing down of the probe pulse. 
For the parameters of KRb molecules, we find that both the Gaussian and hyperbolic secant pulses have a pulse delay of $0.36$ ms  for 1 cm length of propagation.
The group velocity $v_{g}$ calculated from the pulse delay as in Eq.~(\ref{Equation24}) and that from Eq.~(\ref{Equation22}) are in good agreement, and is about $28$ m s$^{-1}$. 
The transmissions through 1 cm are found to be 32.0$\%$ and 34.0$\%$ for the Gaussian and the hyperbolic secant pulses, respectively. 
Further one has to note that the distortion inflicted to the pulse can be minimized by keeping the spectral width of the pulse well inside the transparency window.
 
\begin{figure}[t!]
\includegraphics[width=0.9\columnwidth]{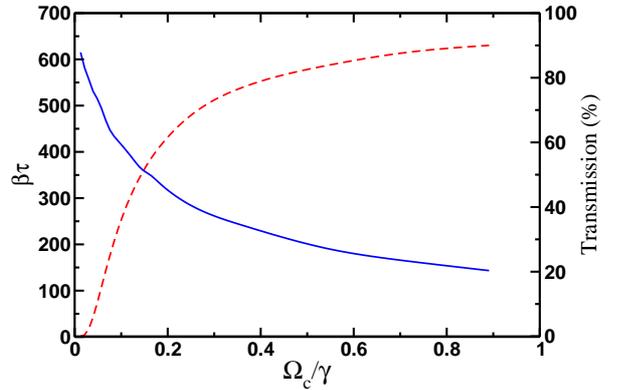}
\caption{\label{fig:Fig7} (Color online) 
Delay-Bandwidth product (solid blue line) and Transmission (red long dashed line) against $\Omega_{c}$ (in unit of $\gamma$) are plotted. Other parameters remain same as in Fig.\ref{fig:Fig3}.}
\end{figure}

Next, we calculate DBP which is a good figure of merit for slow light experiment as mentioned earlier. 
Figure.\ref{fig:Fig7} shows the variation of DBP and transmission as a function of $\Omega_{c}$. 
From this figure we see that as $\Omega_{c}$ increases, DBP decreases. 
This is because of the fact that the rate of decrease in the time-delay of pulse is very high in comparison to the rate of increase of the transparency width. Therefore, the product of time-delay and the spectral width of the transparency window \textit{i.e.}, DBP decreases with the increase in $\Omega_{c}$. 
We can also notice from Fig.\ref{fig:Fig7} that the transmission increases with the increase of $\Omega_{c}$. 
The increase in transmission is due to the increase in the transparency window caused by increasing $\Omega_{c}$. 
Generally, in a slow light experiment, a very low time delay corresponds to a high bandwidth, where as a very high time delay corresponds to a low bandwidth. 
Therefore, one needs to optimize DBP to create a balance between time delay and the bandwidth for efficient slow light applications. 
The optimum DBP in our system is estimated to be around 355, which corresponds to $\Omega_{c}$ of 0.15 $\gamma$, a pulse delay of $0.2$ ms and a transmission of 51.0$\%$ for a Gaussian pulse with a spectral width of 12 KHz.
As mentioned above a main challenge in the slow light experiment is to get a delay equal to the pulse width without any distortion. 
Here we find that a Gaussian pulse of width 0.007 $\gamma$ or equivalently 12 KHz propagating a distance of 1 cm through the medium can exhibit a delay of $0.09$ ms for $\Omega_{c}$ of 0.36 $\gamma $. This amount of delay is nearly equal to the inverse of pulse width. 
Also the pulse is broadened by merely 2$\%$ and the transmission of the pulse is found to be 76.0$\%$. 
  
\section{Conclusion}
In summary, we have studied theoretically a realistic model for pulse propagation in a $\Lambda$-type atom-molecule coupled system. 
We have shown that the width of the transparency window can be controlled by both the stimulated linewidth due to control field as well as Feshbach resonance width. Hence, by controlling the Feshbach resonance width, one can significantly reduce the group velocity of the pulses and also increase the transmittance. We have also shown that a narrow transparency window can lead to a very low group velocity and the spectral width of the pulse should remain well inside the transparency window for distortionless pulse propagation.
Our model can be applied to those ultracold homo- or hetero-nuclear molecules for which the constituents atoms should exhibit narrow Feshbach resonance in ground-state collisions.
We have optimized DBP for an atom-molecule coupled model system for a pulse with spectral width of 12 KHz.
The Optimum DBP for this system is found to be 355.
This work may stimulate further studies into the effects such as optical switching, storage of pulses $\textit{etc.}$ using ultracold atom-molecule coupled systems.  

\end{document}